\documentclass[useAMS,usenatbib]{mn2e}

\def\aap{AA}
\def\apjl{ApJL}
\def\mnras{MNRAS}
\def\apj{ApJ}
\def\aj{AJ}

\def\nat{Nat}

\def\ra{r_{\rm a}}
\def\rt{r_{\rm t}}
\def\nrt{n_{\rm t}}

\def\xa{{\hat r}_{\rm a}}
\def\xt{{\hat r}_{\rm t}}
\def\xtmin{{\hat r}_{\rm t}^{\rm min}}

\def\Rh{R_{\rm h}}
\def\Rht{R_{\rm h,2}}
\def\Rho{R_{\rm h,1}}
\def\Xh{{\hat R}_{\rm h}}
\def\Xho{{\hat R}_{\rm h,1}}
\def\Xht{{\hat R}_{\rm h,2}}

\def\sigmalos{\sigma_{\rm los}}

\def\hatsigma{{{\hat \sigma}}}
\def\sigmalos{{{ \sigma}_{\rm los}}}
\def\sigmaloso{{{ \sigma}_{\rm los,1}}}
\def\sigmalost{{{ \sigma}_{\rm los,2}}}
\def\sigmaloshat{{{\hat \sigma}_{\rm los}}}

\def\msun{{M_\odot}}

\usepackage{graphicx}
\usepackage{float}
\usepackage{amssymb}
\usepackage{amsfonts}
\usepackage{amsmath} 
\usepackage{deluxetable}
\def\Rh{R_{\rm h}}

\title[Multiple Stellar Population in dSphs]{Dark Matter Cores and
  Cusps: The Case of Multiple Stellar Populations in Dwarf Spheroidals}
\author[N. C. Amorisco and
N. W. Evans]{N. C. Amorisco$^{1}$\thanks{E-mail:
    amorisco@ast.cam.ac.uk,nwe@ast.cam.ac.uk} and N. W. Evans$^{1}$\\
  $^{1}$Institute of Astronomy, University of Cambridge, Madingley
  Road, Cambridge CB3 0HA, UK}

\begin{document}

\date{Accepted . Received }

\pagerange{\pageref{firstpage}--\pageref{lastpage}} 

\maketitle

\label{firstpage}

\begin{abstract}
  A number of dwarf spheroidal (dSph) galaxies are known to contain a
  more extended, metal-poor population with a flattish velocity
  dispersion profile, and a more concentrated, metal-rich population
  with a velocity dispersion declining with radius. The two
  populations can be modelled with Michie-King distribution functions
  (DFs) in the isothermal and in the sharply truncated limits,
  respectively. We argue that the truncation of the metal-rich
  population can be traced back to the spatial distribution of the
  star forming gas. Suppose $\delta$ is the exponent of the first
  non-constant term in the Taylor expansion of the total potential at
  the center ($\delta =1$ for Navarro-Frenk-White or NFW halos,
  $\delta =2$ for cored halos). Then, we show that the ratio of the
  half-light radii of the populations $\Rht^{\delta/2}/
  \Rho^{\delta/2}$ must be smaller than the ratio of the line-of-sight
  velocity dispersions $ {{\sigmalost(\Rht)}/{\sigmaloso(\Rho)}}$.

  Specialising to the case of the Sculptor dSph, we develop a
  technique to fit simultaneously both populations with Michie-King
  DFs. This enables us to determine the mass profile of the Sculptor
  dSph with unprecedented accuracy in the radial range $0.2< r < 1.2$
  kpc.  We show that cored halo models are preferred over cusped halo
  models, with a likelihood ratio test rejecting NFW models at any 
  significance level higher than $0.05\%$. Even more worryingly, the best-fit NFW
  models require concentrations with $c\lesssim 20$, which is not in
  the cosmologically preferred range for dwarf galaxies. We conclude 
  that the kinematics of multiple populations in
  dSphs provides a substantial new challenge for theories of galaxy
  formation, with the weight of available evidence strongly against
  dark matter cusps at the centre.
\end{abstract}

\section{Introduction}\label{intro}

The population of dwarf spheroidal satellites (dSphs) of the Galaxy
furnishes an unprecedented opportunity to test the physics underlying
the most extreme dark matter dominated systems. In particular, the
prediction of a cusped density profile for dark matter haloes
\citep{Na96} has naturally focused attention on the structure of the
innermost parts. However, even though reasonable photometric and
kinematic profiles of single stellar populations in several of the
most luminous dSphs are available, the only clear conclusion is that
the methods of stellar dynamics have so far been unable to prove
whether the central regions of such haloes are cored or cusped
\citep[e.g.,][]{Kl02, Ko07, Wa09, Am11}.

The last decade has also seen the growing realization that the stellar
content of dSphs is complex \citep[e.g.,][]{Ma99, Kl04, To04, Ko08,
  Ba08, Ur10}.  Carina, Fornax, Sculptor and Sextans all display
evidence for the co-existence of at least two stellar populations with
different metallicities -- usually an older, more spatially extended
and metal-poor population and a younger, more spatially concentrated,
metal-rich population. In all cases save Carina, the different stellar
populations clearly have different kinematics: while the older
component displays a fairly flattish velocity dispersion profile, the
dispersion profile of the younger component often shows signs of a
quite sharp decline with radius. As a consequence, the metal-rich
population is concentrated in the central regions of the system
only. This picture is shared by both dSphs -- Sculptor and Sextans --
where such kinematic profiles have been completely disentangled
\citep{Ba08, Ba11}.

This is consistent with a picture in which the dSph has undergone
mainly two different star formation epochs, with an underlying
evolution driven by tidal stirring in the potential of the Galaxy.
The metal-poor population may have undergone thermalization with the
help of impulsive tidal stirring near the pericenters of orbits and of
dynamical instabilities driven by the tidal field near the
apocenters \citep[e.g.,][]{Re06, Kl07}. This can lead to a nearly isothermal distribution of
velocities. By contrast, the metal-rich population may originate from
a centrally concentrated distribution of gas, with the gas in the
outskirts of the dSph being removed during the interval between the
two star formation events.  The spatial distribution of the gas may
enforce an approximate cut in energy in the phase space distribution
function of the younger stellar population.

The coexistence of two different stellar populations in the same
potential provides a stronger instrument to probe the properties of
the dark matter halo in dSphs. A first approach is already provided by
\citet{Ba08} in their analysis of the two populations in the Sculptor
dSph. They found that, though a Navarro-Frenk-White (NFW) halo is
still statistically compatible with the kinematical properties of the
two stellar populations of the Sculptor dSph, a cored halo is a better
fit in the context of the Jeans equations. Note however that the
theorem on the central velocity dispersion proved by \citet{An09}
shows that an isotropic cored stellar density profile cannot be
embedded in a cusped dark matter halo, so that Battaglia et al.'s
solutions to the spherical Jeans equations are not all physical 
(see also the recent result provided by \citet{Ci10}). Very
recently, \citet{WaP11} have used an ingenious argument to conclude
that the populations in both the Fornax and Sculptor dSphs strongly
disfavour cusped dark matter haloes. However, their result is based on
the assumption that the mass estimator formula presented in
\citet{Wa09} \citep[see also][]{Wo10} can be applied separately to
both metal-poor and metal-rich stellar populations in a dSph, a
hypothesis which remains to be demonstrated. Although the velocity
dispersion profile of the metal-poor population is consistent with a
quasi-isothermal system, and thus with such a formula, that of the
metal-rich is probably not \citep[c.f.,][]{Ba08}.

At the price of assuming a functional form for the distribution
function of the stellar populations, we can significantly strengthen
the analysis by requiring the models to reproduce at the same time the
luminosity and kinematic profiles for both populations. In doing this
we aim to measure a mass profile (at least in some radial interval),
rather than a single total mass measure at some radius. Both Jeans
analyses \citep[e.g.,][]{Wa09, Wo10} and simple phase space modelling
\citep{Am11} are able to measure the total dark halo mass inside some
specific radius, typically related to the half-light radius of the
stellar population, but have failed to reconstruct the entire mass
profile. This is because, though they are able to link the two
dimensional scales of the dark matter density profile (the
scalelength $r_0$ and the characteristic density $\rho_0$ for
example), they are unable to fix either of these numbers. This is very
typical when the velocity dispersion profile only is fitted, as it is
naturally degenerate in these two dimensional scales.  However, the
surface brightness profile can put independent constraints on the
scalelength of the halo.

We re-analyze the data for the two stellar populations in the Sculptor
dSph by using phase space models, instead of the less complete Jeans
analysis, adopting the general family of Michie-King distribution
functions (King 1962, Michie 1963). This represents a fairly flexible family of centrally
isotropic distribution functions with an adjustable
radial anisotropy in the outer regions. The two stellar populations
are embedded in a dark halo whose density profile can be shaped at
will, cored or cusped.

In Section~\ref{data}, we introduce the observational data; in
Section~\ref{models} we describe the motivations and the basic details
of the models we develop in the paper. Section~\ref{generalconsid}
sets out the properties of our models and establishes a compatibility
criterion between the observed properties of a two populations dSph
and the density profile of the dark matter halo.  Finally, the results
are presented in Section~\ref{results}.

\begin{figure*}
\includegraphics[width=.85\textwidth]{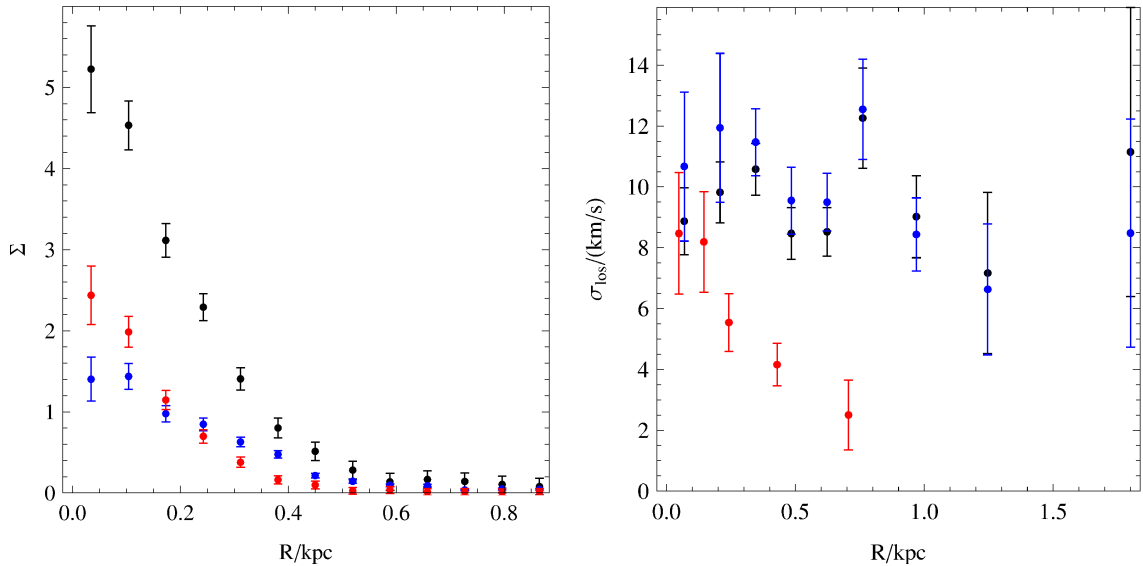}
\caption{Surface brightness profiles (stars arcmin$^{-2}$) and
  projected velocity dispersion profiles for the metal-poor component
  (in blue), metal-rich component (in red) and the red giant branch
  stars (in black). Data from Battaglia et al. (2008a).}
\label{fig:data}
\end{figure*}
%


\section{Sculptor Data}\label{data}

Although a number of our results on multiple populations in dSphs are
general, it is nonetheless useful to have a particular example in
mind. For this, we choose the Sculptor dSph which is located at a
distance $D=79.3$ kpc from the Sun. \citet{Ma99} already provided
evidence for a bimodality in Sculptor's metallicity distribution based
on a properties of the colour-magnitude diagram. This result has been
revised and extended by a number of later investigations
\citep[e.g.,][]{Bab05,  Ba06, Ba08, WaP11}.

\citet{Ba08b} used the FLAMES spectroscope on the VLT in the Ca II
triplet region to obtain high quality radial velocities for Sculptor
members. To divide the stars into metal-rich and metal-poor
populations, \citet{Ba08} assumed a hard cut in metallicity. The
metal-rich population has [Fe/H] $> -1.5$ and traces the red
horizontal branch stars. The metal-poor population has [Fe/H] $< -1.7$
and traces the blue horizontal branch stars. Those stars in the
intervening band of metallicities are not classified. The line of sight velocity dispersion
profiles associated with the red and blue horizontal branch
populations are displayed in Fig.~\ref{fig:data}. Battaglia at
al.'s procedure is open to questioning, as the metallicity is
correlated with the kinematics, and the procedure may introduce biases
that are not easy to quantify. We explored this issue by reanalizing the 
entire radial velocity detaset, looking for correlations between
the position of the hard cut in metallicity and the derived kinematics.
We find that the kinematics of the metal-poor population is robust against 
changes in the cut position, as well as against foreground contamination
at large radii (different ways to estimate the number of contaminants -- see 
eqn~(1) in \citet{Ba08b} -- do not result in significant variations of the 
velocity disperion profile). The metal-rich population is more prone to
variations, though we find that binning size is more of an issue when compared
to the position of the metallicity cut itself, which leaves the main features
of the velocity dispersion profile unchanged if varied within $0.1$ in [Fe/H].



\citet{Ba08} also provide surface brightness profiles for the red and
blue horizontal branch stars, which are also shown in
Fig.~\ref{fig:data}.  A photometric profile is usually characterised
by the half-light radius, namely the (projected) radius the encloses
half the light. Battaglia et al. find the metal-poor population is
well fit by a Plummer profile with half-light radius $15.1 \pm 0.5$
arcmin ($348\pm 12$ pc), whilst the metal-rich prefers a slightly
cuspier Sersic profile with half-light radius $8.6 \pm 1$ arcmin
($198\pm 25$ pc).  We re-compute the two half light radii directly
from the surface brightness profiles in Fig.~\ref{fig:data} using a 
model-independent technique, which does not require the assumption of any 
precise functional form for the surface brightness profiles themselves.
We assume a Gaussian distribution for the observational uncertainties 
of the surface brightness profiles and use them to generate syntetic surface brightness
profiles through a Monte Carlo technique. By direct integration of these profiles
we infer the half-light radii of the two stellar populations as well as their uncertainties 
(variations due to the order of the interpolation of the syntetic profiles are accounted 
for in the final uncertainty itself). We obtain:
\begin{equation}
\Rho  =  341 \pm 9\ {\rm pc},\qquad
\Rht  =  217 \pm 15\ {\rm pc},
\label{hlr}
\end{equation}
for the half-light radii of the metal-poor and metal-rich populations
respectively.

\section{Models}\label{models}

\subsection{The Dark Halo}

Both stellar populations are assumed to be embedded in a massive,
extended, spherical dark matter halo, the density profiles of which
can be shaped at will. The dark halo is truncated at the tidal radius
of the metal-poor stellar population. Given the high mass-to-light
ratios of dSphs, we assume that the entire gravitational field is
determined by the dark matter only and that the stellar components are
just tracers.

As a flexible family of dark matter haloes, we use
\begin{equation}
\rho(r)= \begin{cases} {\rho_0 \over
\left({r \over r_0}\right)^a
 \left(1+\left({r\over r_0}\right)^b\right)^{c/b}} & r < \rt,\\
0 & r > \rt. \end{cases}
\label{eq:genNFW}
\end{equation}
Here, $r_0$ and $\rho_0$ are the dimensional scalelength and
characteristic density of the halo. In our models, the tidal radius
$\rt$ of the dark matter halo is the same as that of the metal-poor
stellar population. Even if, in nature, the truncation radius of the
dark matter halo is larger than the stellar tidal radius, the
potential experienced by the stars remains unchanged. The metal-rich
population is allowed to have a smaller tidal radius.

Throughout the paper, we will concentrate on the following two
choices:
\begin{equation}
\begin{array}{ccc}
(a, b, c) & = & (1, 1, 2)\\
(a, b, c) & = & (0, 2, 3)\\
\end{array}
\label{eq:3profs}
\end{equation}
The first choice yields the cosmologically-motivated
Navarro-Frenk-White (NFW) model. The second is a standard example of
cored model, with the same asymptotic density fall-off as the NFW
profile ($\rho \sim r^{-3}$).

\subsection{The Stellar Populations}\label{MKdf}

We are inspired by the fact that, as a result of relaxation processes,
the final distribution function of the stars in phase space tends to
near-isothermality, at least in the very central parts. This idea is
supported by the universal flatness of the velocity dispersion
profiles of the dSphs at least close to the centres (Evans, An, Walker
2009), as well as numerical simulations of tidal stirring of dSphs
(Mayer et al. 2001). Fig.~1 shows that for the Sculptor dSph the
metal-poor population has a flat dispersion profile out to at least
$\sim 500$ pc.  The metal-rich profile is perhaps more debatable,
though consistent with a flattish profile within $\sim 200$ pc.

Deviations from flatness of the velocity dispersion profiles can be
ascribed to either anisotropy or tidal effects. In principle, since
the two populations reside in the same potential and experience the
same tidal perturbation, they should share the same cut in energy too
if this is tidal in origin. However, we are motivated by the idea that
the metal-rich population has indeed been formed by gas which was
concentrated in the central and more protected regions of the dSph.
Thus, its cut-off in radius may be much smaller than the formal tidal
radius. In what follows, we therefore assume that the more extended
metal-poor population has the same tidal radius as the dark halo, but
allow the radial truncation of the metal-rich population to be set by
the data.

The Michie-King distribution function or DF (King 1962, Michie 1963)
is a function of two integrals of the motion, the energy $E$ (per unit
mass) and the angular momentum $L$ (per unit mass), namely
\begin{equation}
f_{\rm MK}(E,L)={\rho_{*,0}\over{(2\pi
\sigma^2)^{3/2}}}\exp\left({{-L^2}\over{2\ra^2\sigma^2}}\right)f_{\rm K}(E)\
,
\label{eq:michieking}
\end{equation}
where 
\begin{equation}
f_{\rm K}(E)=\exp \left[{{\Phi(\rt)-E }\over{\sigma^2}}\right]-1.
\label{eq:king}
\end{equation}
The gravitational potential $\Phi$ is, in general, the total
gravitational potential of the system, but here we assume that the
stars can be treated as simple tracers.

The dimensional free parameters of the Michie-King DF are as follows:

\medskip\noindent (1) the velocity dispersion parameter $\sigma$,
which, for a fixed gravitational potential, tunes the scalelength of
the associated luminous component;

\medskip\noindent (2) the tidal radius $\rt$, which is the external
edge of the system, truncated by the tidal effects of the Galaxy, or
by the original energy distribution of the star forming gas;

\medskip\noindent (3) the anisotropy radius $\ra$, which is
approximately the characteristic length scale associated with
variations in the anisotropy profile;
 
\medskip\noindent 
(4) the characteristic stellar density $\rho_{*,0}$. As long as the
stars are just tracers in the potential of the dark halo, this
parameter is an overall normalization, and has no other effect on the
model.

\medskip To our set of dimensional parameters, we can associate a
useful set of dimensionless parameters. We use the scalelength of the
halo $r_0$ and its characteristic density $\rho_0$ as dimensional
references, and denote dimensionless quantities by a super-script hat,
so that the dimensionless half-light radius, anisotropy radius and
tidal radius are $\Xh = \Rh/r_0$, $\xa=\ra/r_0$ and $\xt=\rt/r_0$
respectively.  The dimensionless velocity dispersion is
\begin{equation}
\hatsigma^2\equiv {\sigma^2\over \Phi_0}\  \  \  ;\  \  \  \sigmaloshat^2(\hat r)\equiv {\sigmalos^2(r/r_0)\over \Phi_0} \ ,
\label{eq:dimlesss}
\end{equation}
where, respectively for an NFW and a cored halo, we have (see also \citet{Am11})
\begin{equation}
\Phi_0^{\rm NFW}=2\pi G \rho_0 r_0^2, \qquad
\Phi_0^{\rm core}=2\pi G \rho_0 r_0^2/3.
\label{phi0def}
\end{equation}
A Michie-King model for a single stellar population in a dSph is
defined by the three numbers $(\hatsigma, \xt, \xa)$ and two scalings
set by the dark halo $(r_0, \rho_0)$. A two-population model for a
dSph therefore has $3+3+2 =8$ free parameters.

\begin{figure}
\includegraphics[width=.48\textwidth]{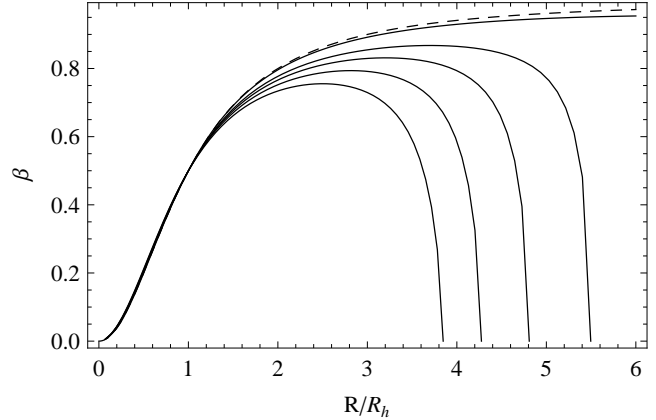}
\caption{Anisotropy profiles of Michie-King models embedded in an NFW
  halo, all having $\Xh=0.2$ and $\bar{\beta}=1/2$, but with different
  truncations. The dashed profile shows the Osipkov-Merrit
  limit~(\ref{OMbeta}).  }
\label{aniprof}
\end{figure}

Michie-King DFs are never tangentially anisotropic, and range from
perfectly isotropic to highly radial. This does not seem unduly
restrictive on examining the velocity dispersion profiles in Fig.~1,
both of which decline with radius. This fact does not suggest any
great preponderance of tangential bias in the orbits.  The anisotropy
profiles $\beta(r)=1-\sigma^2_{\rm t}/(2\sigma^2_r)$, (where
$\sigma^2_{\rm t}=\sigma^2_{\theta}+\sigma^2_{\phi}$ is the tangential
velocity dispersion) of Michie-King DFs are always isotropic at the
centre, but can be adjusted at larger radii through the anisotropy
radius $\xa$. However, Michie-King models with the same anisotropy
radius may have completely different degrees of anisotropy (depending
on the values of the other two free parameters). Thus, in the
following, we prefer to parametrize the anisotropy of our models using
the value of the anisotropy profile at the half-light radius
\begin{equation}
{\bar {\beta}}=\beta(\Rh) .
\label{betarh}
\end{equation}
Also, in order to avoid systems subject to radial orbit instability (e.g.,
Barnes, Hut \& Goodman 1986), we only consider $\bar{\beta}\leq1/2$.

\subsection{The Isothermal and the Strongly Truncated Limits}

To understand the properties of Michie-King models, it is helpful to
distinguish two opposite physical regimes. The DFs can vary from
exponential in the energy, typical of isothermal systems, to a
power-law dependence in the `reduced` energy $E-\Phi(\rt)$, typical of
strongly truncated systems. Notice that we are exploring a much wider
range of Michie-King models than is usual in their most common
application, the study of globular clusters.

In the isothermal limit, most of the stars are unaffected by the tidal
cut in King`s DF~(\ref{eq:king}), which is thus dominated by the
exponential term:
\begin{equation}
{{\left[\Phi(\rt)-\Phi(r)\right]}/{\sigma^2}}\gg 1\ .
\label{eq:ineq}
\end{equation}
An equivalent way to state this inequality is $\Xh \ll \xt$.  When
this holds true, the dependence of the Michie-King DFs on energy $E$
and angular momentum $L$ becomes close to the linear combination
$E+L^2/(2\ra^2)$, which describes the Osipkov-Merritt case (Osipkov
1979, Merritt 1985).  Thus, in this instance, the anisotropy
asymptotes to the profile
\begin{equation}
  \beta_{\rm OM}(r) \sim {r^2\over{r^2+r_a^2}}\ .
\label{OMbeta}
\end{equation}
This describes systems which are isotropic in the center ($r\ll\ra$)
and become more radially anisotropic at larger radii ($r\gg\ra$). The
isotropic, and therefore fully isothermal limit, is equivalent to
additionally imposing
\begin{equation}
\Xh / \xa  \ll  1 .
\label{eq:ineqs1}
\end{equation}
This corresponds to the case in which the half-light radius of the
stars is smaller than the other length-scales in the problem. In this
regime, isothermality ensures a flat velocity dispersion
profile. Furthermore, if the stars are deeply embedded in their dark
matter halo (i.e. $\Xh\ll 1$), then the stellar profile has a Gaussian
density-law if embedded in a cored halo, and an exponential one if
embedded in an NFW halo. Amorisco \& Evans (2011) show that these
systems satisfy:
\begin{equation}
  \Rh/r_0 \propto (\sigma^2/\Phi_0)^{1/\delta} \approx (\sigmalos^2/\Phi_0)^{1/\delta}\ ,
\label{centralbeh}
\end{equation}
where $\delta$ is the exponent of the first non-constant term in the
Taylor expansion of the total potential at the center. So, $\delta =1$
for the NFW case, $\delta = 2$ for a cored dark halo.

The opposite of the isothermal regime is the case in which the
exponential term in King`s DF~(\ref{eq:king}) is of order unity. This
limit describes strongly truncated systems, in which the dimensionless
half-light radius $\Xh$ and the dimensionless tidal radius $\xt$ are
of the same order. The asymptotic expression for the anisotropy is:
\begin{equation}
  \beta_{\rm trunc}(r)\sim{2\over9} {r^2\over{\ra^2}} {{\Phi(\rt)-\Phi(r) }\over{\sigma^2}}\ .
  \label{truncbeta}
\end{equation}
This is true for any Michie-King model close to its tidal edge.  So,
any Michie-King model becomes isotropic on approaching its tidal
radius. This is evident from the anisotropy profiles plotted in
Fig.~\ref{aniprof}.

Strongly truncated isotropic models are essentially scale-free models,
whose properties depend only on the gravitational potential $\Phi(r)$
and the tidal radius $\rt$. The radial velocity dispersion is
\begin{equation}
\sigma^2_r(r) = {2\over 7} (\Phi(\rt)-\Phi(r))\ ,
\label{plummerdens}
\end{equation}
which generates cut-offs in the line of sight velocity dispersion
profiles. For the stellar density distribution, we have
\begin{equation}
\rho_*(r) \propto (\Phi(\rt)-\Phi(r))^{5/2}\ .
\label{plummerdens}
\end{equation}

For fixed half-light radius $\Xh$ and degree of radial anisotropy, the
tidal radius realized by strongly truncated systems is the smallest in
the entire family of Michie-King models with a given gravitational
potential.  Thus, we use this limit to define the function
\begin{equation}
\xtmin(\Xh, \bar{\beta})\equiv \lim_{\hatsigma\to \infty} 
\xt(\hatsigma,\Xh, \bar{\beta}) \ .
\label{xtmin}
\end{equation}
It is often convenient to express the tidal radius as multiples of
thi minimum value defined by eqn.~(\ref{xtmin}), that is
\begin{equation}
{\nrt}\equiv {\xt \over \xtmin} \geq 1\ ,
\label{nrt}
\end{equation}
where $\nrt$ is the tidal number. This is useful in presenting our
results, as $1/\nrt$ lies between 0 and 1.

The Michie-King model for a single population embedded in a dark halo
can be described by the natural set of 3 parameters, namely
dimensionless velocity dispersion $\hatsigma$, tidal radius $\xt$ and
anisotropy radius $\xa$. An alternative, but equivalent set, is the
dimensionless half-light radius $\Xh$, the tidal number $\nrt$ and the
anisotropy parameter at the half-light radius $\bar{\beta}$. It is
this set that we will use in the rest of the paper.

\begin{figure}
\includegraphics[width=.95\columnwidth]{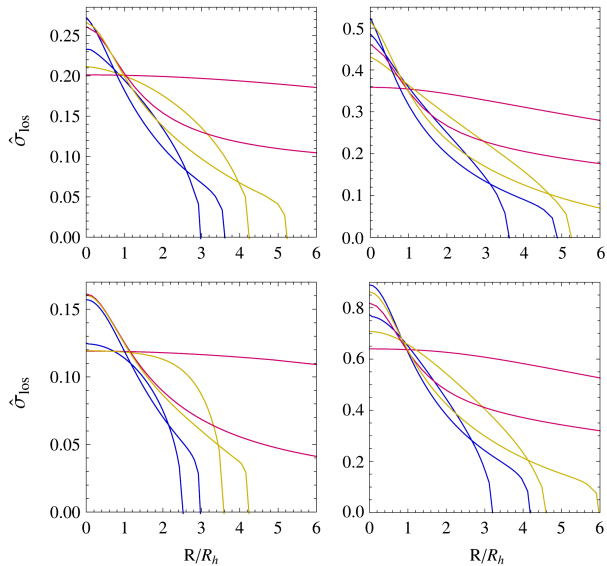}
\caption{The line of sight velocity dispersion profiles generated by
  Michie-King DFs embedded in a NFW dark matter profile (upper panels)
  and in a cored dark matter halo (lower panels). From left to right,
  $\Xh=0.1$, $\Xh=1$. In each panel, profiles of different colours
  have different tidal radii, beginning with the smallest possible
  $\xtmin (\Xh, \bar{\beta})$, in blue. The two profiles with the same
  colour in each panel have different radial anisotropies:
  $\bar{\beta}=0$ and $\bar{\beta}=0.5$, at the extremes of our
  domain. }
\label{profilesrangenfw}
\end{figure}
\begin{table}
 \centering
  \begin{tabular}{@{}c|ccc@{}}
    \hline
         & $\Xh=0.1$ & $\Xh=0.3$ & $\Xh=1$  \\
    \hline
 $\sigmaloshat(\Xh)$ & 1.6\% & 1.5\% & 1.7\% \\
 $\sqrt{\langle \sigmalos(R)^2 \rangle_{\Sigma(R)}}$ & 4.0\% & 6.1\% & 4.5\% \\
   \hline
   $\sigmaloshat(\Xh)$ & 2.0\% & 1.5\% & 1.2\% \\
 $\sqrt{\langle \sigmalos(R)^2 \rangle_{\Sigma(R)}}$ & 5.9\% & 4.6\% & 3.7\% \\
    \hline
\end{tabular} 
\caption{The relative uncertainty obtained using the luminosity weighted
  average or the value at $\Xh$ in the mass estimator. In each
  instance, the upper line refers to NFW haloes, the lower to cored
  haloes.}
\label{discrep}
\end{table} 
%

%
\begin{figure}
\centering
\includegraphics[width=.48\textwidth]{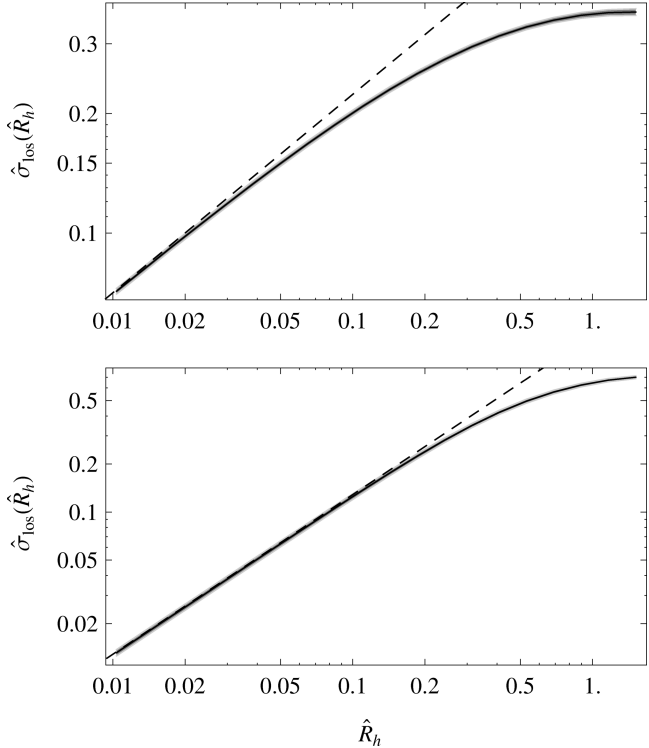}
\caption{The relation (with scatter) between the dimensionless half
  light radius $\Xh$ and dimensionless line of sight velocity
  dispersion at $\Xh$, $\sigmaloshat(\Xh)$ for Michie-King models
  embedded in an NFW halo (upper panel) and in cored halo (lower
  panel). In both panels, the dashed line represents the asymptotic
  analytic relation provided by the isothermal limit. }
\label{relssigma1}
\end{figure}

\section{Properties of the models}\label{generalconsid}

\subsection{Dark Halo Mass Estimators}\label{sigmarel}

A main aim in studying stellar populations in dSphs is to map the
underlying dark halo. A number of authors have recently suggested that
the mass of the dark halo within a characteristic radius related to
the half-light radius is well-constrained (e.g., Walker et al. 2009,
Wolf et al. 2010, Amorisco \& Evans 2011). These mass estimators all
take the general form (c.f. Illingworth 1976)
\begin{equation}
  M(\lambda \Rh) = {K \over G}  \Rh \langle \sigma_{\rm los}^2 \rangle
\label{eq:illing}
\end{equation}
where $K$ and $\lambda$ are two dimensionless constants and $\langle
\sigma_{\rm los}^2 \rangle$ is an appropriate `average' of the square
of the line of sight velocity dispersion. A successful mass estimator
needs to fix each of these three ingredients.

Let us begin with the average $\langle \sigma_{\rm los}^2 \rangle$.
In the isothermal limit, the profile is almost flat by definition, so
we do not need to worry at which radius we pick the line of sight
velocity dispersion. When the dispersion profile falls with radius, it
is a moot point as to how best to provide the `average'.  This topic
assumes some importance as it is the heart of the argument provided by
Walker \& Penarrubia (2011) in their analysis.

To bring the problem of the `average` into sharp focus, consider two
stellar systems embedded in the same dark matter halo and with the
same half-light radius, but characterized by a different DF in phase
space, and thus having different kinematics and density profiles. For
example, consider a radially-biased and a tangentially-biased
system. A proper recipe for an efficient mass estimator should pick
the same `average' velocity dispersion for both systems, though they
will in general differ both in their velocity dispersion profile
$\sigmalos(R)$ and their surface brightness profile $\Sigma(R)$.  Wolf
et al. (2010) suggested that a good choice is the luminosity-weighted
average line of sight velocity dispersion ${\langle \sigmalos(R)^2
  \rangle_{\Sigma(R)}}$.  However, a radially-biased model will have
both higher velocity dispersion profile and higher surface brightness
profile in the central regions and lower in the outer regions, as
compared to the tangential model, which militates against the
usefulness of a luminosity-weighted average.

Fig.~\ref{profilesrangenfw} illustrates the dimensionless line of
sight velocity dispersion profiles for models that span the entire
physical range of Michie-King models, for NFW and cored halos
(upper and lower panels respectively). The panels show models with
different half-light radii, tidal radii and anisotropy. In each panel,
all the models have the same dimensionless half-light radius $\Xh$ and
should then provide the same mass estimate and `average`. A quick look
makes it evident that the easiest choice is the line of sight velocity
dispersion at one $\Xh$, where all the profiles roughly cross. The
effectiveness of this choice is readily quantified. For three
different values of the dimensionless half-light radius $\Xh=0.1$,
$\Xh=0.3$ and $\Xh=1$, Table~\ref{discrep} collates the relative
uncertainty obtained by comparing the line of sight velocity
dispersion at one half light radius $\sigmaloshat(\Xh)$ of all
possible Michie-King models ($0 < \bar{\beta} < 0.5$, $0 < 1/\nrt <
1$). The same uncertainty is calculated for the luminosity averaged
velocity dispersion $\sqrt{\langle \sigmalos(R)^2
  \rangle_{\Sigma(R)}}$. Our recipe always performs better, with the
luminosity-weighted average $\sqrt{\langle \sigmalos(R)^2
  \rangle_{\Sigma(R)}}$ usually showing discrepancies higher than
$4\%$.

Fig.~\ref{relssigma1} displays the functional relations obtained by
associating each dimensionless half-light radius $\Xh$ to the
dimensionless line of sight velocity dispersion $\sigmaloshat(\Xh)$
for an NFW halo (upper panel) and for a cored halo (lower panel). In
both panels the dashed line represents eqn.~(\ref{centralbeh}), which
is valid in the isothermal limit for $\Rh\ll 1$. The displayed shaded
areas correspond to the error bars as in Table~1, calculated by
comparing the line of sight velocity dispersions $\sigmaloshat(\Xh)$
for models with different $\bar{\beta}$ and $\nrt$, but identical
half-light radius $\Xh$. These relations are similar to the ones displayed 
in Fig.~2 of \citet{Am11}, which were derived for isotropic King models in the 
isothermal regime only.

The functional relations $\sigmaloshat(\Xh)$ allow us to calculate
directly the two dimensionless quantities $\lambda$ and $K$ which
define the mass estimator~(\ref{eq:illing}).  We find
\begin{equation}
\lambda =1.67\pm0.04, \qquad
K =5.85\pm0.2.
\label{eq:ineqs}
\end{equation}
This result is in agreement with \citet{Am11}, but it is worth noting
its much wider range of validity, which covers the entire physical
range of Michie-King models. Summarizing, we obtain
\begin{equation}
  M\left[(1.67\pm0.04)\Rh\right] = {(5.85\pm0.2 )}  {{\Rh \sigma_{\rm los}(\Rh)^2}\over G}\ ,
\label{illingfin}
\end{equation}
which can be considered trustworthy even when radial anisotropy and
substantial deviations from isothermality may be present.

\begin{figure}
\centering
\includegraphics[width=.48\textwidth]{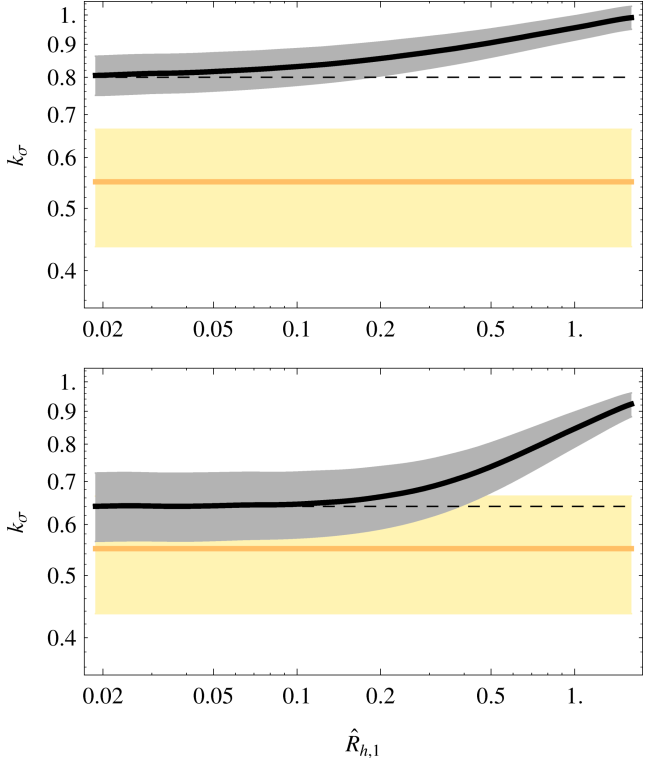}
\caption{In black, with uncertainty range in gray: the ratio
  $k_{\sigma}\equiv\sigmaloshat(\Xht)/\sigmaloshat(\Xho)$ against
  $\Xho$ when the ratio $\Rht/\Rho$ is assumed from the available
  data. Upper panel: NFW halo; lower panel: cored halo. In both panels
  the yellow shaded area represents the poor evidence for $k_{\sigma}$
  provided by the observed line of sight velocity dispersion profiles;
  the dashed line represents the asymptotic value provided by the
  isothermal limit.}
\label{compatib}
\end{figure}
%

\subsection{A Consistency Criterion for Two Populations}

As the two stellar populations reside in the same dark matter halo,
their dimensionless half light radius $\Xh$ and line of sight velocity
dispersion $\sigmaloshat(\Xh)$ must be picked from the same functional
relation in Fig.~\ref{relssigma1}. This means that the ratios of the
half-light radii and the velocity dispersion
\begin{equation}
k_R \equiv {\Rht \over \Rho}, \qquad\qquad
k_{\sigma}\equiv {{\sigmalost(\Rht)}\over{\sigmaloso(\Rho)}},
\label{ksigmadef}
\end{equation}
are not independent quantities. They are related to one another
through the density profile of the dark matter halo itself. In
general, we must have
\begin{equation}
\left( {\Rht \over \Rho} \right)^\delta \leq
\left( {{\sigmalost(\Rht)}\over{\sigmaloso(\Rho)}} \right)^2
\implies k_R^{\delta}  \leq k_\sigma^2
 , 
\label{criterion}
\end{equation}
with equality attained in the isothermal limit. This can be understood
from Fig.~4, as the ratio $\hat R_{{\rm h},
  i}^\delta/\sigmaloshat(\hat R_{{\rm h}, i})^2$ is an increasing
function of $\hat R_{{\rm h}, i}$ and attains unity in the isothermal
limit for strongly embedded systems.

Let us now specialise to the case of Sculptor.  For Battaglia's
two-population dissection, using eqn~(1), we find that 
\begin{equation}
k_R =0.64\pm0.05\ .
\end{equation}
This measure (with its uncertainty) and the relations in
Fig.~\ref{relssigma1} (with their uncertainties) generate the black
line (and gray shaded areas) in Fig.~\ref{compatib}. All the pairs
$(\Xho, k_{\sigma})$ in the gray areas are compatible with an NFW
halo (upper panel) or with a cored halo (lower panel) for some value
of the halo`s scalelength $r_0$.  Also displayed in the figure as the
asymptotic dashed line is the isothermal limit.

Measurement of the ratio $k_{\sigma}$ remains a challenge despite
substantial improvements in the quality of the photometric and
spectroscopic data in recent years.  We interpolate both the velocity
dispersion profiles, and the profiles defined by the uncertainties,
and obtain
\begin{equation}
k_{\sigma}= 0.55\pm0.12  \ ,
\end{equation}
which is displayed in Fig.~\ref{compatib} as the yellow shaded area.

It is evident that the cored halo provides a much better
interpretation of the available data, as there is a formal agreement
between models and data which is not the case for the NFW
halo. Furthermore, we can extrapolate this result and state that any
other cusped dark matter density profile with a cusp steeper than the
NFW prescription is going to deliver even worse results.

In passing, we note that our consistency criterion~(\ref{criterion})
can be reformulated to provide the one used by \citet{WaP11}. In fact,
if two populations fail to satisfy the inequality~(\ref{criterion})
for a given $\delta$, then the logarithmic `slope` $\Gamma$, estimated
using eqn~(\ref{eq:illing}), will satisfy the inequality
\begin{equation}
  \Gamma={{\ln\left[M(\lambda\Rht)/M(\lambda\Rho)\right]}\over{\ln\left(\Rht/\Rho\right)}}=1+2{{\ln k_{\sigma}}\over{\ln k_R}}\geq \delta+1\ ,
\label{criterion2}
\end{equation}
thus confirming that such a system is inconsistent with the density
profile $\rho\sim r^{\delta-2}$.

\begin{figure}
\centering
\includegraphics[ width=.95\columnwidth]{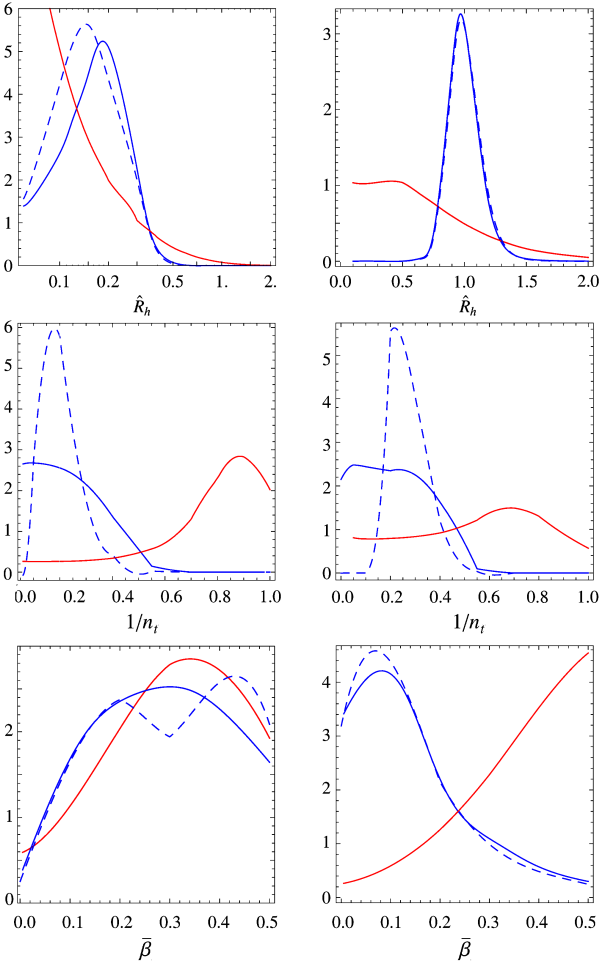}
\caption{The probability distributions associated with the likelihoods
  $L_{\rm MP}$ and $L_{\rm MR}$ for the free parameters of the model
  when embedded in a NFW (left) and cored (right) halo. The blue (red)
  profiles are associated with the metal-poor (metal-rich) stellar
  component. Dashed and full blue profiles show the results deriving
  from the total likelihood~(\ref{Ltot1}) and the `reduced`
  likelihood~(\ref{Ltot1nort}). }
\label{fig:projcomb}
\end{figure}
%

\section{Phase space analysis: Sculptor}\label{results}

To perform a complete maximum likelihood analysis of the Sculptor data
using Michie-King DFs necessitates an exploration of an
eight-dimensional parameter space. In fact, the main body of this
analysis can be successfully reduced to the parallel and independent
studies of the two metal-poor and metal-rich stellar populations,
ignoring at first that they reside in the same dark matter halo.  The
combination of the evidence coming from the two stellar populations
then constrains the range of feasible dark matter haloes.

The details of the fitting of the Michie-King models to the
kinematics, the photometry and -- in the case of the metal-poor
population -- the tidal radius are given in Appendix A. Here, we note
that the fit to the metal-poor population yields a $\chi^2_{\Sigma}$,
$\chi^2_{\sigma}$ and $\chi^2_{\rt}$, whilst the metal-rich yields has
only a $\chi^2_{\Sigma}$ and $\chi^2_{\sigma}$. This is because the
tidal radius of the metal-rich population is free to vary, whilst that
for the more extended metal-poor must be the same as the underlying
dark halo.

The combination of all the available evidence for the metal-poor
stellar population is provided by the likelihood and associated
$\chi^2$-squared values:
\begin{eqnarray}
L_{\rm MP}(\Xh, \nrt, \bar{\beta}, r_0, \rho_0) &=&L_{\Sigma} \cdot L_{\sigma} \cdot L_{\rt}\ \nonumber\\
\chi_{\rm MP} & =&\chi^2_{\Sigma} + \chi^2_{\sigma} + \chi^2_{\rt}.
\label{Ltot1}
\end{eqnarray}
However, for the sake of clarity, we will also consider the reduced
likelihood
\begin{equation}
L_{\rm MP}^{\rm red}(\Xh, \nrt, \bar{\beta}, r_0, \rho_0) =L_{\Sigma} \cdot L_{\sigma} \ ,
\label{Ltot1nort}
\end{equation}
partly to check that the fitting of the tidal radius does not have an
undue and distorting effect on our results.  Similarly, for the
metal-rich population, we can define
\begin{eqnarray}
L_{\rm MR}(\Xh, \nrt, \bar{\beta}, r_0, \rho_0) &=& L_{\Sigma} \cdot L_{\sigma} \ ,\nonumber\\
\chi_{\rm MR} &=& \chi^2_{\Sigma} + \chi^2_{\sigma}\ .
\label{Ltot2}
\end{eqnarray}

\begin{table}
 \centering
  \begin{tabular}{@{}c|ccc@{}}
    \hline
         & $\chi^2_{\Sigma}$ & $\chi^2_{\Sigma}+\chi^2_{\sigma}$ & $\chi^2_{\Sigma}+\chi^2_{\sigma}+\chi^2_{\rt}$\\
    \hline
    NFW & 39.3, 41.5, 45.7 & 48.2, 66.9, 68.3 &  49.0, 67.8, 69.7 \\
        & 1.97, 2.08, 2.29 & 1.12, 1.56, 1.59 &  1.14, 1.58, 1.62 \\
    
   \hline
    cored & 32.7, 36.3, 39.7 & 39.0, 54.3, 59.6 & 40.3, 55.7, 60.5 \\
          & 1.63, 1.81, 1.98 & 0.91, 1.26, 1.39 & 0.94, 1.29, 1.41 \\
    
    \hline\hline

    NFW & 2.3, 5.6, 9.1 & 3.4, 11.1, 13.8 &  - \\
        & 0.38, 0.93, 1.52 & 0.34, 1.11, 1.38 &  - \\
    
   \hline
    cored & 1.1, 3.3, 4.9 &  2.1, 6.8, 9.9 &  -  \\
          & 0.18, 0.55, 0.82 &  0.21, 0.68, 0.99 &  -  \\

    \hline\hline
  \end{tabular} 
  \caption{Results of the independent analysis of the metal-poor
    (upper) and metal-rich (lower) stellar component. The table gives 
    the values of the $\chi^2$-quantities referring, in order, to the 
    best fit models, to the 68\% and to the 95\%
    confidence regions, which have been calculated
    using the likelihoods in eqns~(\ref{LSigma}, \ref{Lsigma},
    \ref{Lrt}). Also the corresponding reduced-$\chi^2$-quantities 
    are reported in the second line for each halo model.
    Uniform priors have been assumed in the range
    defined by $0 < \Xh < 2$, $0 \le 1/\nrt < 1$, $0 < \bar{\beta} < 0.5$,
    as well as for $r_0$ and $\log\rho_0$.}
\label{resnonsgrav1}
\end{table} 

\subsection{Independent Analysis of the Two Populations}
\label{sepnongrav}

Table~\ref{resnonsgrav1} collects the results pertaining the
metal-poor and metal-rich stellar populations when fitted separately,
thus allowing each of them to have different dimensional scales $r_0$
and $\rho_0$.  For each of ($\chi^2_{\Sigma}$,
$\chi^2_{\Sigma}+\chi^2_{\sigma}$,
$\chi^2_{\Sigma}+\chi^2_{\sigma}+\chi^2_{\rt}$), the values associated
respectively with the best fit model, the 68\% and the 95\% confidence
regions are displayed.

A comparison between the values for an NFW and cored halo makes clear
that the main difference is in the reproduction of the stellar surface
brightness distribution, with the cored halo performing substantially
better.  Also, the fact that the increase in $\chi^2_{\sigma}$ is very
similar for an NFW and cored halo shows that the shape of the velocity
dispersion profile is largely insensitive to the dark halo properties.

For the metal-poor (metal-rich) population, Fig.~\ref{fig:projcomb}
displays in blue (red) the probability distributions of the free
parameters of the Michie-King models, as deduced from the projection
of the total likelihood.  First, while the cored halo selects a
Michie-King model with a similar scale length ($\Xh\approx1$), an NFW
halo needs to be several times more extended ($\Xh\lesssim 0.2$) than
the distribution of stars in order to fit (yet with difficulties) the
surface brightness profile of the metal-poor stars.  Also, in the
cored case, an almost isotropic metal-poor stellar population is
preferred ($\bar{\beta}\approx 0.1$), whilst an NFW halo favours a
mild radial velocity bias ($\bar{\beta}\approx 0.25$). The metal-rich
population requires a high degree of radial anisotropy
($\bar{\beta}\gtrsim 0.4$), irrespective of choice of halo. Note also
the necessity of different tidal radii in the phase space
distributions is supported by the results. In both cases,
$\nrt\lesssim 1.4$, suggesting that the metal-rich stellar population
is indeed in a different physical regime compared to the metal-poor
one, which has $n_{\rm t}\gtrsim 3$. Finally, the probability distributions for the dimensionless
half-light radius $\Xh$ of the metal-rich population show that little
help in determining the characteristic radius of the halo can be
expected from a population in this heavily truncated regime.

\begin{figure}
\includegraphics[width=.43\textwidth]{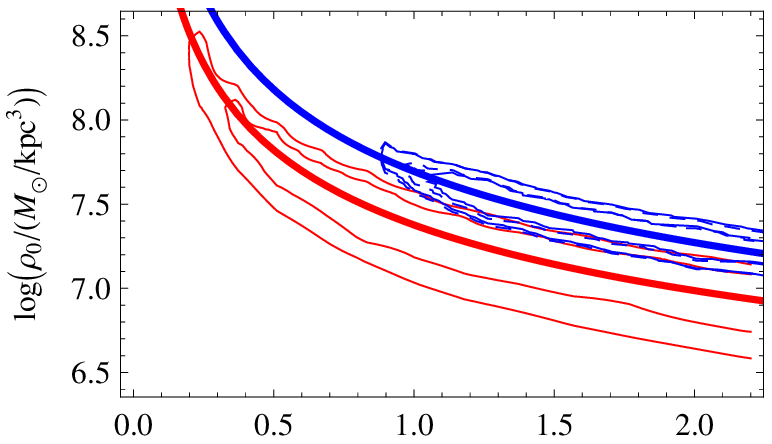}
\includegraphics[width=.43\textwidth]{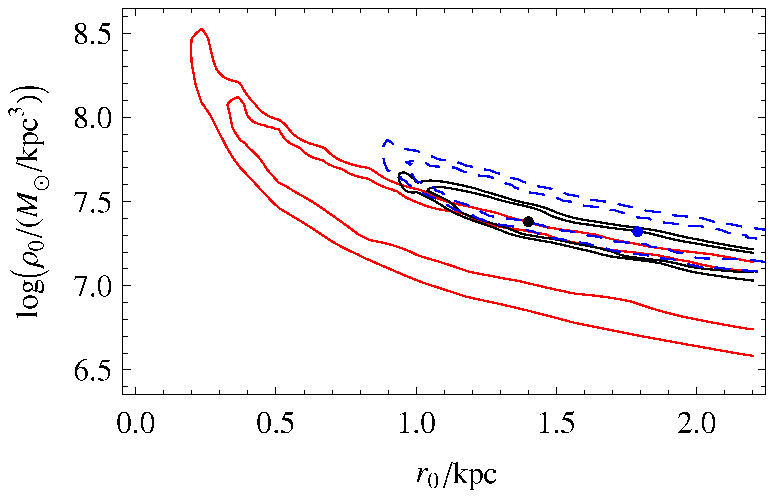}
\caption{The 68\% and 95\% confidence regions in the $(r_0, \rho_0)$
  plane for an NFW halo. The blue (red) contours are associated with
  the likelihood $L_{\rm MP}$ ($L_{\rm MR}$) for the metal-poor
  (metal-rich) stellar population.  Dashed and full blue profiles show
  respectively the results deriving from the total
  likelihood~(\ref{Ltot1}) and the reduced
  likelihood~(\ref{Ltot1nort}).  The black lines display the contours
  of the joint evidence. Full dots indicate the best fit models.}
\label{fig:likplanenfw}
\end{figure}
\begin{figure}
\includegraphics[width=.43\textwidth]{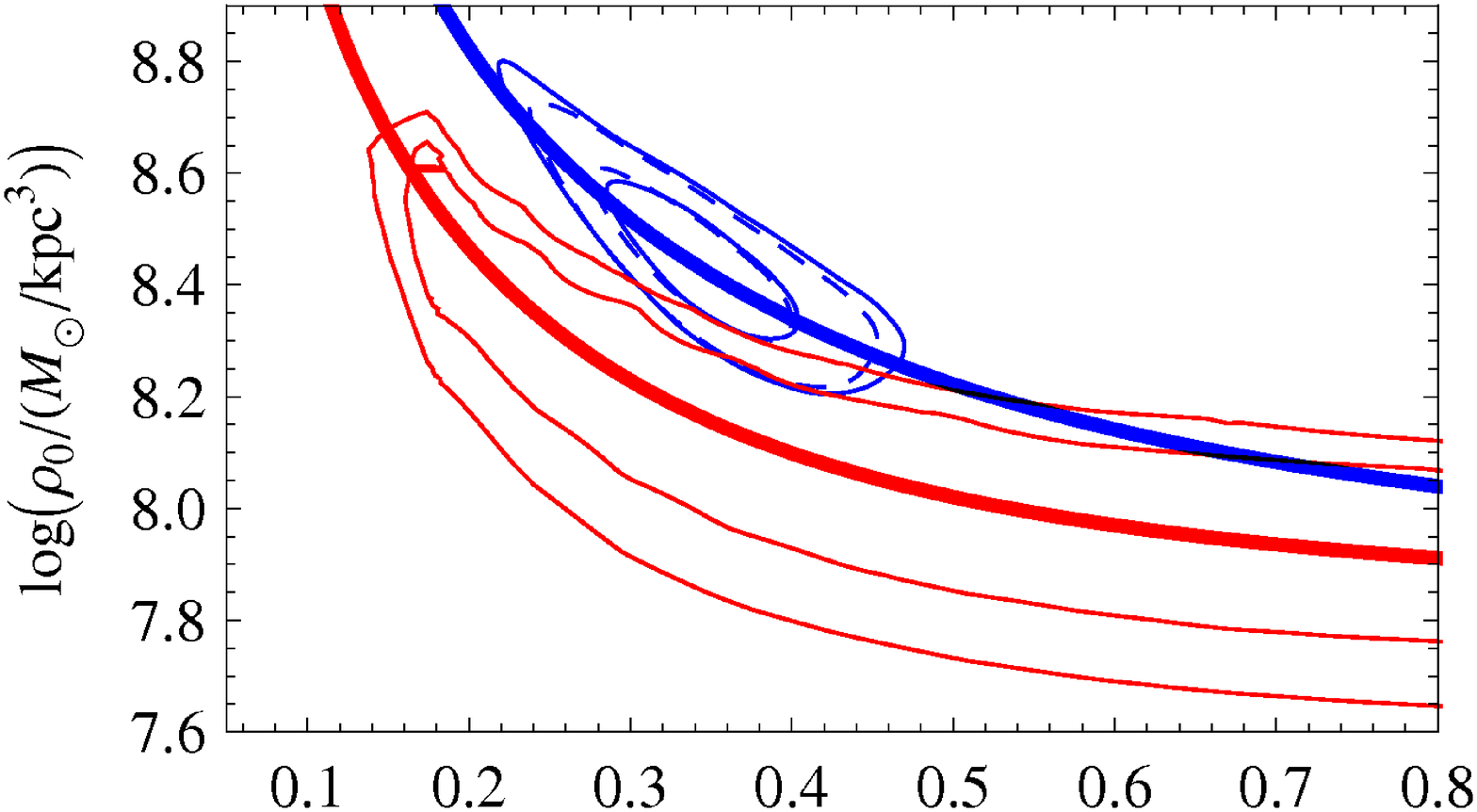}
\includegraphics[width=.43\textwidth]{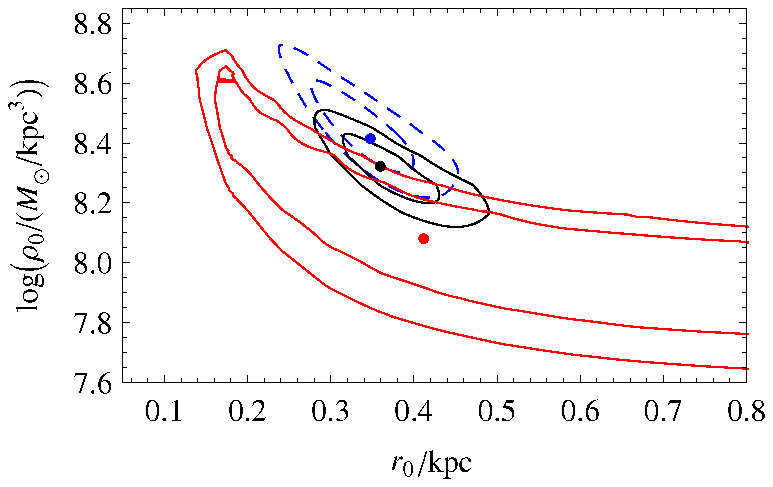}
\caption{As Fig.~\ref{fig:likplanenfw}, but for the cored halo.}
\label{fig:likplanecore}
\end{figure}
\begin{figure*}
\includegraphics[width=.7\textwidth]{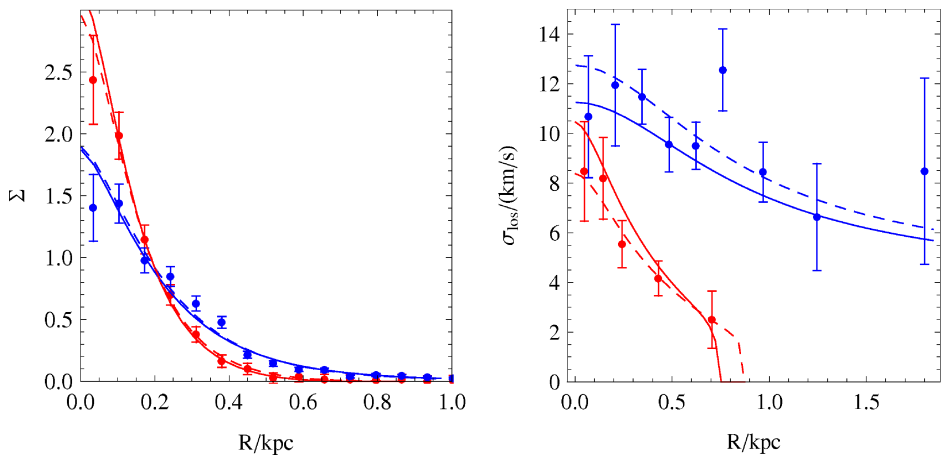}
\caption{The photometric and kinematic profiles for the two Sculptor
  populations in an NFW halo. Thick profiles: final fits provided by
  the best fitting Michie-King models associated with the black dot in
  Fig.~\ref{fig:likplanenfw}. Dashed profiles: best fits provided by
  the Michie-King models, but when $(r_0, \rho_0)$ is allowed to vary
  separately for the two stellar components. Surface brightness
  profiles are measured in stars arcmin$^{-2}$.}
\label{fig:finalprofsnfw}
\end{figure*}
\begin{figure*}
\includegraphics[width=.7\textwidth]{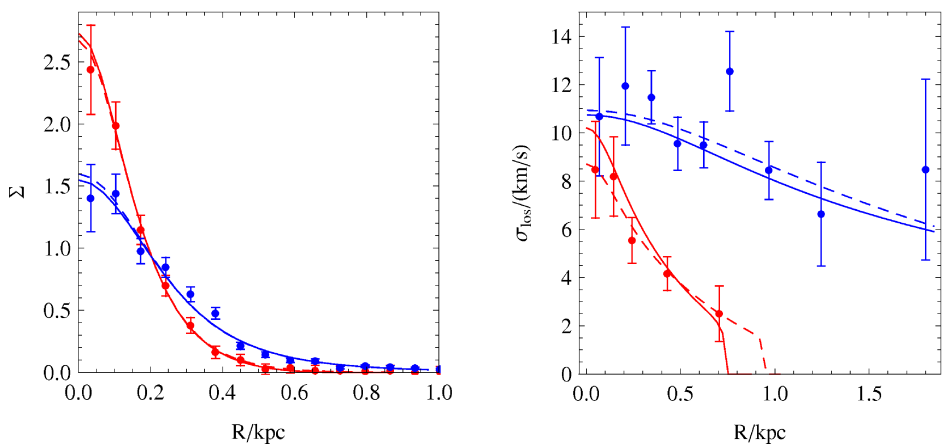}
\caption{As Fig.~\ref{fig:finalprofsnfw}, but for a cored halo.}
\label{fig:finalprofscore}
\end{figure*}
%

\subsection{Two Populations in the Same Halo}\label{combanal}

Figs~\ref{fig:likplanenfw} and~\ref{fig:likplanecore} display the
projections of our likelihoods onto the halo plane $(r_0, \rho_0)$,
for an NFW and a cored halo respectively.  In both Figs, the upper
panel shows the 68 \% and 95 \% confidence regions associated with the
separate analysis of the two stellar populations with blue (red)
representing the metal-poor (metal-rich) respectively.  The full thick
blue and red lines in the upper panels of both figures show the trends
of the $\rho_0(r_0)$ functional relations obtained respectively for
the metal-poor and metal-rich stellar populations by using the
relations $\sigmaloshat(\Rh)$ displayed in Fig.~\ref{relssigma1} for
an NFW and a cored halo.

The lower panel shows how the independent results from the two stellar
populations combine. We in fact impose the condition that the
metal-poor and metal-rich stellar population live in the same dark
matter halo $(r_0, \rho_0)$ via:
\begin{eqnarray}
  L_{\rm tot}(m_1, m_2, r_0, \rho_0) &=&L_{\rm MP}(m_1, r_0, \rho_0) \cdot L_{\rm MR}(m_2, r_0, \rho_0)  \nonumber\\
  \chi_{\rm tot} & = &\chi^2_{\rm MP} + \chi^2_{\rm MR} \ ,
\label{Ltot}
\end{eqnarray}
where $m_i=(\Xh, \nrt, \bar{\beta})$ is a general Michie-King model.
Full dots indicate the best fit models.

Finally, Table 3 collects, for an NFW and cored halo, the values of
$\chi_{\rm tot}$ associated with the best fitting models, the details
of which we report here. For an NFW halo, we find:
\begin{eqnarray}
  (r_0, \rho_0)&=&(1.4{\rm \ kpc}, 2.3 \times 10^7 
  \msun{\rm \ kpc}^{-3}),\nonumber\\
  m_1&=&(\Xh=0.256, \nrt = 10.0, \bar{\beta} =0.20),\nonumber\\
  m_2&=&(\Xh=0.136, \nrt = 1.1, \bar{\beta} = 0.37)
  \ .
\label{nfwbf}
\end{eqnarray}
In physical units, the Michie-King DF for the metal-poor population
has a tidal radius of $r_{\rm t}$ of 12.2 kpc, an anisotropy radius
$r_{\rm a}$ of 720 pc and a velocity dispersion $\sigma$ of 11.0
kms$^{-1}$. The same quantities for the metal-rich population are
$r_{\rm t}$ of 750 pc, $r_{\rm a}$ of 140 pc and a velocity dispersion
$\sigma$ of 15.6 kms$^{-1}$. For a cored halo, we find:
\begin{eqnarray}
  (r_0, \rho_0)&=&(0.36{\rm \ kpc}, 2.1 \times 10^{8} \msun{\rm \ kpc}^{-3}),\nonumber\\
  m_1&=&(\Xh=0.98, \nrt=6.7, \bar{\beta}=0.1),\nonumber\\
  m_2&=&(\Xh=0.54, \nrt=1.1, \bar{\beta}=0.48)
  \ .
\label{corebf}
\end{eqnarray}
This means that the Michie-King DF for the metal-poor population has
$r_{\rm t} = 7.8$ kpc, $r_{\rm a} = 1.0$ kpc and $\sigma_0 = 10.8$
kms$^{-2}$. The metal-rich has $r_{\rm t} = 750$ pc, $r_a = 120$ pc
and $\sigma = 16.0$ kms$^{-2}$.

There are a few points that it is worth noting. As remarked already,
neither the velocity dispersion profile nor the surface brightness
profile are able to constrain the characteristic scalelength of the
NFW dark halo. This is reinforced by Fig.~\ref{fig:likplanenfw}, in
which only the lower limit $r_0^{\rm NFW}\gtrsim 1$ kpc is fixed. The
contours of the confidence regions associated with $L_{\rm tot}$ do
not close as $r_0$ becomes large.  The contours do follow accurately
the functional relation $\rho_0^{\rm NFW}(r_0)$ described by the blue
or red thick full curves.  By contrast, the confidence regions for the
metal-poor populations do close for a cored halo. This is a
consequence of the surface brightness profile, which picks out a
limited interval in the scalelength $r_0$, thus constraining at the same
time the central density $\rho_0$. The confidence regions for the
metal-rich populations still do not close, though this is a
characteristic of the strongly truncated limit of the Michie-King
models.

In both Figs~\ref{fig:likplanenfw} and~\ref{fig:likplanecore}, only
marginal agreement can be found between the 68\% confidence region
associated with the metal-poor and the metal-rich stellar
populations. However, the importance of this is minimal in the case of
a cored halo, since it does not cause any significant increase in the
area of the confidence regions (or in the chi-square values)
associated with the joint likelihood. Note, in fact, that the values
of total $\chi^2_{\rm tot}$ associated with the joint evidence $L_{\rm
  tot}$, do prefer a dark matter density profile with a constant
density core.

Following \citet{Ea71}, we define a set of models with nested
parameters, by considering halos with a ${1-\lambda}$ cusp and a
density fall of $\rho(r)\sim r^{-3}$ at large radii. This obviously
includes the cusped NFW density profile ($\lambda =0$) and the cored
profile ($\lambda = 1$). Since the difference in best-fit total
$\chi^2$ values between the NFW case and the cored case is
$58.5-46.3=12.2$, we can reject a pure NFW density profile at any
significance level higher than $0.05\%$. Completely analogous results
have been obtained when the original kinematic profiles taken from Battaglia et al. (2008)
have been replaced with kinematic profiles derived directly from the radial velocity dataset
using slightly different metallicity cuts (within 0.1 [Fe/H]) or membership criteria (see also discussion in Sect.~2).
Furthermore, although not reported here in
detail, we investigated the issue of a milder cusp by considering
separately the $\lambda =1/2$ case. A milder cusp performs better than
the NFW halo, but we can still reject it in favour of a cored halo at
any significance level higher than $1.4\%$ (we find a best-fit total
$\chi^2$ of 52.3). All available indications suggest a monotonic
behaviour of the best-fit total $\chi^2$ with the cusp index.

Furthermore, the best fitting NFW model~(\ref{nfwbf}), together with
the parts of the halo $(r_0, \rho_0)$ plane supported by the joint
evidence $L_{\rm tot}$, sit in a region in discord with the standard
expectation of cosmological models. The concentration of the best
fitting NFW halo is in fact as low as $c\approx 17$, whereas
present-day dwarf galaxies are expected to have higher concentrations, 
$c \approx 35$ according to Colin et al. 2004 (but see also \citet{Ma07}).

Figs~\ref{fig:finalprofsnfw} and~\ref{fig:finalprofscore} show the
quality of the final fits to the surface brightness profiles and
velocity dispersion profiles respectively in the case of an NFW halo
and a cored halo. The full profiles are the best joint fits, that is
obtained by using the same $(r_0, \rho_0)$ halo for both the
metal-poor and the metal-rich stellar population. The dashed profiles
show instead the best separate fits. It is evident from these plots
that the real damage to the prospects of fitting the data with an NFW
halo come from two sources: it is the combination of photometry and kinematics that does
the damage. First, although the properties of the
metal-poor population are reproduced reasonably well, the metal-rich
population is predicted to have a higher central surface density than
the data. Second, the NFW halo requires a smaller difference than the
observed one between the values of the metal-poor and metal-rich
velocity dispersion in order to reproduce the observed half-light
radius. This smaller difference is a rephrasing of the inconsistence 
found in Fig.~5 through our consistency criterion~(\ref{criterion}). 
However, note that the addition of the information about the 
surface brightness profiles of the two stellar populations -- 
which was not included in the arguments presented in Section 4 --
has in fact changed the range of best fitting cored models. While the consistency 
criterion alone would select haloes with a larger scalelength ($\Xh\lessapprox 0.5$, see Fig.~5),
the complete maximum likelihood analysis prefers models in which, approximately, mass follows light ($\Xh\approx 1$).

\begin{table}
 \centering
  \begin{tabular}{@{}c|cc@{}}
    \hline
    & $\chi^2_{\rm tot, bf}$ & $\chi^2_{\rm tot}$ \\
    \hline
    NFW     & 58.5 &  78.1, 80.2  \\
    cored    &  46.3 &  62.2, 68.0  \\
    \hline
\end{tabular} 
\caption{The values of $\chi^2_{\rm tot}$ associated with the best 
  fitting halo, as well as the 68\% and 95\% confidence regions, 
  determined according to the likelihood~(\ref{Ltot}).}
\label{jointtable}
\end{table} 

\section{The Mass Profile for Sculptor}

\begin{figure}
\includegraphics[width=.48\textwidth]{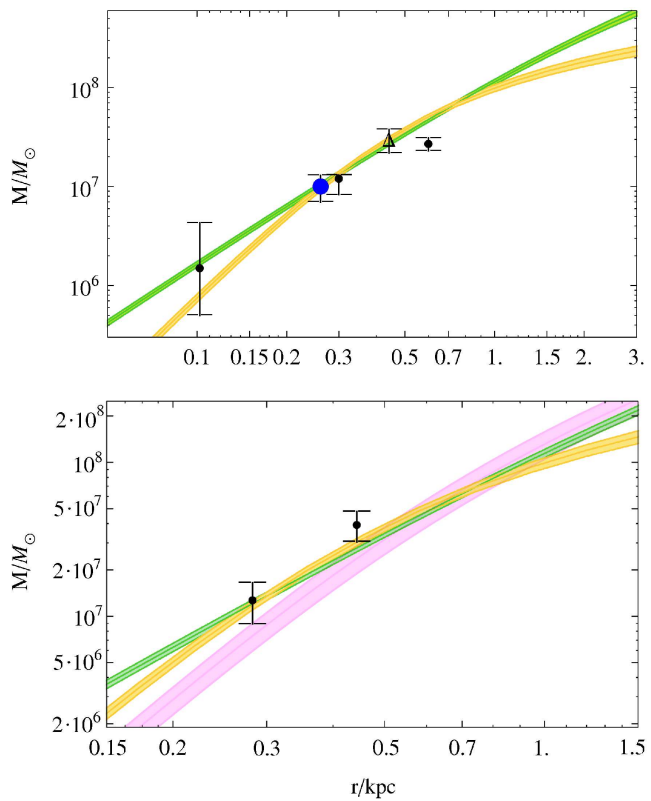}\\
\caption{Upper panel:The mass profiles with errors as deduced from
  Figs.~7 and 8 for the NFW halo (in green) and the cored halo (in
  yellow). In the top panel, Strigari et al.'s (2007, 2009) mass
  measures at 100 pc, 300 pc and 600 pc are shown as black dots,
  Walker et al.'s (2009) measure at $\Rh=260$ pc is shown as a blue
  dot and Amorisco \& Evans (2011) measurement at $1.7\Rh\approx 442$
  pc is shown as a triangle. Lower panel: A zoom from the top panel
  displaying the agreement in the range $0.2 \lesssim r \lesssim 1.2$
  kpc, together with the masses estimated using Wolf et al.`s (2010)
  formula.}
\label{massprofiles}
\end{figure}

Fig.~\ref{massprofiles} shows the final mass profiles for the Sculptor
dSph. In both upper and lower panel, the coloured shaded areas
represent, at each radius, the 1-$\sigma$ uncertainty obtained for the
mass profile using the joint evidence $L_{\rm tot}$, for the case of
an NFW halo (in green) and for a cored halo (in yellow).

In the cored case, we expect to be able to measure a mass profile
rather than a single mass enclosed within some radius. This is because
we are able to fix the best $(r_0, \rho_0)$ in the halo plane. What is
more surprising is that the same thing is true for the NFW halo, even
though we have only a lower limit for the characteristic radius
$r_0$. However, this can be understood by noticing that, as
anticipated in Amorisco \& Evans (2011), the mass profile $M(r,r_0)$
generated by the functional relation $\rho_0(r_0)$ becomes insensitive
to the characteristic radius $r_0$ when $\Xh\ll 1$.

The points displayed in the upper panel show how our results compare
with previous mass measures for Sculptor by Strigari et al. (2007,
2008), Walker et al. (2009) and Amorisco \& Evans (2011). The lower
panel is a zoom for a smaller range of radii. We draw attention to the
fact that, even if the matter of a core or cusp at the centre is
considered unresolved, we are still able to measure a {\it mass
  profile} rather than the single enclosed mass datapoint at a
particular radius. The mass profiles pertaining to the cored and NFW
halo, in fact, agree within their 1-$\sigma$ uncertainty over the
entire interval $200{\rm pc}\lesssim r \lesssim 1.2$kpc, thus proving
a wide and unprecedented radial coverage. The pink shaded area
represents the best-fit cored halo as reported in~\citet{Ba08}.

Also displayed in the lower panel of Fig.~\ref{massprofiles} are the
two mass estimates obtained using the formula suggested by Wolf et
al. (2010), and applied separately to the metal-poor and metal-rich
stellar populations. We calculate each of the two luminosity-averaged
velocity dispersions $\sqrt{\langle \sigmalos(R)^2
  \rangle_{\Sigma(R)}}$ using the observed profiles and a Monte Carlo
technique which assumes Gaussian uncertainties. Note that the
logarithmic slope of the mass profile we would deduce from such a
technique is steeper than the one we actually measure.

\section{Conclusions}

The suggestion that stellar kinematics in dwarf spheroidals (dSphs)
may enable the structure of the dark matter halo to be mapped out is
at least a decade old. It motivated programs to gather radial
velocities in the bright dSphs with 4m and 8m class telescope and
multi-object spectrographs. Despite a lot of work, it has proved hard
to distinguish the structure of the dark matter halo, especially at
small radii, where cold dark matter (CDM) theories predict $1/r$
density cusps (e.g., Navarro, Frenk \& White 1996).
 
For example, Gilmore et al. (2007) provided a succinct summary of the
then available observational evidence.  They noted that analyses based
on the Jeans equations for six dSphs (Draco, Ursa Minor, Carina,
Sextans, Leo I and Leo II) tended to suggest that the dark matter
density was shallower than the $1/r$ density cusp. However, they
acknowledged that the observed velocity dispersion profiles and cored
light distributions were likely to be consistent with dark matter
halos with both central cores and cusps.

A thorough examination of the evidence using the Jeans equations for
the eight brightest dSphs (the six studied by Gilmore et al. together
with Fornax and Sculptor) was carried out by Walker et al. (2009)
using datasets of thousands of stars.  They found that the most
meaningful constraint provided by a Jeans analysis was on the total
mass enclosed within the half-light radius (see also Wolf et al. 2010).
In particular, the Jeans equations were not sufficiently powerful to
distinguish between cored and cusped dark matter haloes.  A companion
theoretical investigation by Evans, An \& Walker (2009) made explicit
the limitations of the Jeans equations. They showed how assumptions as
to the surface brightness and profile velocity anisotropy could allow the
misleading identification of a core or a cusp. Nonetheless, this is a
consequence of the interlocking assumptions made in the Jeans
analysis, and not the data, which are consistent with both cores and
cusps.

However, in at least two dSphs, there appears to be direct evidence
favouring dark halos with constant density cores. First, Kleyna et
al. (2003) discovered kinematically cold substructure in the Ursa
Minor dSph. They argued from numerical simulations that the persistence
of such cold substructure is incompatible with the cusped profiles of
cold dark matter cosmogonies. They provided an interpretation in terms
of a stellar cluster dissolving in a cored dark matter halo.  Second,
Goerdt et al. (2006) and \citet{Sa06} studied the five globular clusters present in the
Fornax dSph, which orbit at projected radii $\sim 1$ kpc from the
centre. In a cusped dark matter halo, they would sink to the centre
under the action of dynamical friction in a few Gyr. In a cored dark
matter halo, the dynamical friction timescale is much longer and so
this provides a natural resolution of the survival problem of Fornax's
globular clusters.

The exploration of multiple stellar populations in dwarf spheroidals
is only just beginning. But there are now clear hints that it might
provide a resolution of this problem. The fact that both populations
reside in the same dark matter halo significantly reduces the
degeneracies.  To date, there have been three different attempts at
modelling -- each with their own strengths and limitations. It is
striking that all three have come to the conclusion that the data are
much more consistent with cored dark matter haloes than cusped one.

Battaglia et al.'s (2008) investigation of the case of Sculptor used
the Jeans equations and came to the conclusion that the data are
statistically consistent with both cored and cusped haloes, though the
NFW profile yields a poorer fit for the metal-rich stars.  The
drawback to this analysis is that it is based on the Jeans equations
alone, and there is no guarantee that the underlying models exist. In
fact, the central velocity dispersion proved by \citet{An09} shows
that an isotropic cored stellar density profile cannot be embedded in
an Navarro-Frenk-White halo, and so there is no solution for the
phase-space distribution function for the metal-poor population,
despite the existence of the Jeans solution. Nonetheless, Battaglia et
al.'s (2008) basic conclusion -- that a cored halo model provides a
better fit than an NFW halo model -- seems robust and is recovered by
our analysis.

Walker \& Pe{\~n}arrubia (2011) developed a sophisticated statistical
method for assigning probabilities as to whether dSph stars belong to
metal-rich and metal-poor sub-populations. This is a significant
improvement on the simple cut in metallicity used by Battaglia et
al. (2008), which can introduce biases when metallicity is correlated
with kinematics.  As output from their statistical analysis, they
obtain a measure of the average velocity dispersion of the metal-poor
and metal-rich populations. This is used with a mass estimator of the
form ~(\ref{eq:illing}) to obtain the mass enclosed at two points,
namely the half-light radius of the metal-poor and the metal-rich
populations. From this, the gradient of the dark halo mass profile is
calculated. For the Sculptor and Fornax dSphs, they conclude that
cusped NFW profiles are ruled out. The modelling is then simpler than
a Jeans analysis, but Walker \& Pe{\~n}arrubia (2011) argue that it is
more robust. One qualm is that the method assumes that the two
components are drawn from constant velocity dispersion populations,
both in the statistical analysis itself and in the application of the
mass estimator formula. Whilst this seems reasonable enough for the
metal-poor populations in dSphs, it is not at all clear that the
metal-rich population can be so described. In fact, the combined or
total velocity dispersion of the total population must then decline
outwards, which is certainly not the case in, at least,
Fornax. Nonetheless, Walker \& Pe{\~n}arrubia have tested their
reasoning against synthetic datasets generated from a wide range of
phase-space distribution functions, and it performs very well.

Finally, in this paper, we have used the Sculptor dataset provided by
Battaglia et al. (2008) to build phase-space distribution functions
for both the metal-poor and the metal-rich populations under the
assumption of cored and cusped dark matter haloes. Although both can
fit the data, cored dark matter haloes are preferred. There is clear
evidence of discrepancies in the surface brightness profile of the
metal-rich stars near the centre for cusped haloes. Notice that it is
the combination of surface photometry with the stellar kinematics --
rather than the kinematics alone -- that is providing the decisive
evidence against cusped haloes. Even more worryingly, the
Navarro-Frenk-White models that are the best-fit are much less
concentrated than the cosmological predictions from cold dark matter
theories. Typically, concentrations of $c \approx 35$ are predicted
from simulations of galaxy formation (e.g., Colin et al. 2004), as
compared against $c \approx 17$ found in our analysis.

The fact that all three analyses -- which make different assumptions
and have different strengths and weaknesses -- come to the same
conclusion is very striking.  Further modelling of other dSphs with
multiple populations is needed to confirm the results. However, the
evidence that the dark haloes of the dSphs around the Milky Way do not
have cusped form, but have cores, is now beginning to look rather
strong.  Understanding the origin of these constant density dark
matter cores could be one of the most exciting challenges facing
astronomy in the next few years.

\section*{Acknowledgments}
We thank
Giuseppina Battaglia, Mike Irwin and Eline Tolstoy for kindly making
their data on Sculptor available to us.  Discussions with Matthew
Walker and Jorge Pe{\~n}arrubia have been very thought-provoking and
interesting. We thank them for making available their manuscript prior
to publication.  It is a pleasure to thank the referee, Mark
Wilkinson, for his helpful comments and to acknowledge valuable
suggestions from Giuseppe Bertin and Jan Conrad. NA thanks STFC and
the Isaac Newton Trust for financial support.

\appendix

\section{Full phase space analysis:  the technique}\label{nonselfgrav}

\subsection{The surface brightness profile}\label{sbrighanal}

For any population defined by a Michie-King model $(\Xh, \nrt,
\bar{\beta})$ in a dark halo characteristic radius $r_0$, we can
define the quantity
\begin{equation}
  \chi^2_{\Sigma}=\sum_i {{\left[\Sigma^m_0 \hat\Sigma^m(R_i/r_0; \Xh, \nrt, \bar{\beta})-\Sigma^o(R_i)\right]^2}\over{{\rm Var}[\Sigma^o(R_i)]}}\ . 
\label{chiSigma}
\end{equation}
Here, the collection of points $(R_i, \Sigma^o(R_i))$ is the observed
surface brightness profile, each of the points having its variance
${\rm Var}[\Sigma^o(R_i)]$.  The function $\hat\Sigma^m(\hat{r}; \Xh,
\nrt, \bar{\beta})$ is the dimensionless surface brightness of the
Michie-King model with parameters $(\Xh, \nrt, \bar{\beta})$ at the
dimensionless point $\hat{r}$ (measured in units of the halo
scalelength).  This surface brightness is made comparable with the
data by the dimensional coefficient $\Sigma^m_0$, which is determined
in such a way to give the smallest $\chi^2_{\Sigma}$:
\begin{equation}
{{\partial\chi^2_{\Sigma}}/{\partial \Sigma^m_0}} = 0\ .
\label{chiSigmaconds}
\end{equation}
We associate the quantity $\chi^2_{\Sigma}$ to the standard likelihood
\begin{equation}
L_{\Sigma}(\Xh, \nrt, \bar{\beta}, r_0)=\exp\left(-\chi^2_{\Sigma}/2\right)\ .
\label{LSigma}
\end{equation}
%

\subsection{The velocity dispersion profile}\label{vdispanal}

Similarly, for any triple $(\Xh, \nrt, \bar{\beta})$ and any pair of
dimensional scales $r_0$, $\rho_0$, we define the quantity
\begin{equation}
\chi^2_{\sigma}=\sum_i {{\left[\sqrt{\Phi_0}\ \sigmaloshat^m(R_i/r_0; \Xh, \nrt, \bar{\beta})-\sigmalos^o(R_i)\right]^2}\over{{\rm Var}[\sigmalos^o(R_i)]}}\ .
\label{chisigma}
\end{equation}
Here, the collection of the points $(R_i, \sigmalos^o(R_i))$ is the
observed projected velocity dispersion profile, each of the points
having its variance ${\rm Var}[\sigmalos^o(R_i)]$. The function
$\sigmaloshat^m(\hat{r}; \Xh, \nrt, \bar{\beta})$ is the dimensionless
projected velocity dispersion profile of the Michie-King model with
parameters $(\Xh, \nrt, \bar{\beta})$ at the dimensionless point
$\hat{r}$ (measured in units of the halo scalelength). This velocity
dispersion is made comparable with the data by the square root of the
potential term $\Phi_0$, eqn~(\ref{phi0def}), and in which the
dimensional scales $r_0$ and $\rho_0$ enter.  We associate the
quantity $\chi^2_{\sigma}$ to the standard likelihood
\begin{equation}
L_{\sigma}(\Xh, \nrt, \bar{\beta}, r_0 ,\rho_0)=\exp\left(-\chi^2_{\sigma}/2\right)\ .
\label{Lsigma}
\end{equation}
%

\subsection{The tidal radius of the metal-poor stellar
  component}\label{rtanal}

As explained in Section~\ref{MKdf}, the two stellar components are
allowed to have different tidal radii. Recall that we interpret the
tidal radius of the metal-rich as an approximate cut in energy of its
phase space distribution. On the other hand, the tidal radius of the
metal-poor stellar component is the true tidal radius of the dSph
within the gravitational potential of the Galaxy.  At this radius, the
density distribution of the halo is truncated too, according to the
prescription of eqn.~(\ref{eq:genNFW}). Hence, such a tidal radius has
to satisfy, at least approximately, the Roche criterion:
\begin{equation} {{M_{\rm dSph}(\rt)}\over{\rt^3}} = {{M_{\rm
        MW}(D-\rt)}\over{(D-\rt)^3}}\ ,
\label{Rochec} 
\end{equation}
where $M_{\rm dSph}(r)$ is the mass profile of the dwarf, $M_{\rm
  MW}(r)$ is the mass profile of the Galaxy and $D$ is their
separation.

Suppose we have fixed a Michie-King model $(\Xh, \nrt, \bar{\beta})$
and the pair of dimensional scales $r_0$, $\rho_0$.  Then the tidal
radius of the model is
\begin{equation}
  \rt^{m}(\Xh, \nrt, \bar{\beta}, r_0) = \nrt\cdot\xtmin(\Xh, \bar{\beta})\cdot r_0\ .
\label{rtmodel} 
\end{equation}
We denote this by $\rt^R(\Xh, \nrt, \bar{\beta}, r_0 ,\rho_0)$ and
quantify the discrepancy using
\begin{equation}
\chi^2_{\rt}={{\left(\rt^m-\rt^R\right)^2}\over{\left(\delta\rt^R\right)^2}}\ .
\label{chirt}
\end{equation}
This compares the two different determinations of the tidal radius
$\rt^m$ and $\rt^R$.  The uncertainty $\delta\rt^R$ takes into account
the uncertainty on the mass profile of the Galaxy, for which we allow
the interval $2\gtrsim M_{MW}/(10^{12}M_{\odot})\gtrsim 0.5$.  We
finally associate the quantity $\chi^2_{\rt}$ to the standard
likelihood via
\begin{equation}
L_{\rt}(\Xh, \nrt, \bar{\beta}, r_0 ,\rho_0)=\exp\left(-\chi^2_{\rt}/2\right)\ .
\label{Lrt}
\end{equation}

\end{document}